\documentclass[aps,prl,amsmath,amssymb,superscriptaddress,showkeys,showpacs,twocolumn,floatfix]{revtex4-1}

\usepackage{graphicx}
\usepackage{subfigure}
\usepackage{hyperref}

\usepackage{xcolor}
\usepackage[utf8]{inputenc}

\begin{document}


\title{Cosmic QCD Epoch at Nonvanishing Lepton Asymmetry}

\author{Mandy M. Wygas}
\email[]{m.wygas@physik.uni-bielefeld.de}
\affiliation{Fakult\"at f\"ur Physik, Universit\"at Bielefeld, Postfach 100131, 33501 Bielefeld, Germany}
\author{Isabel M. Oldengott}
\email[]{isabel.oldengott@uv.es}
\affiliation{Departament de Fisica Te\`orica, Universitat de Val\`encia, 46100 Burjassot, Spain}
\author{Dietrich B\"odeker}
\email[]{bodeker@physik.uni-bielefeld.de}
\affiliation{Fakult\"at f\"ur Physik, Universit\"at Bielefeld, Postfach 100131, 33501 Bielefeld, Germany}
\author{Dominik J. Schwarz}
\email[]{dschwarz@physik.uni-bielefeld.de}
\affiliation{Fakult\"at f\"ur Physik, Universit\"at Bielefeld, Postfach 100131, 33501 Bielefeld, Germany}

\date{\today}

\begin{abstract}
We investigate how a lepton asymmetry impacts the cosmic trajectory in the QCD phase diagram. 
We study the evolution of chemical potentials during the QCD epoch of the early Universe
using susceptibilities from lattice QCD to interpolate between an ideal quark gas and an ideal 
hadron resonance gas. The lepton asymmetry affects the evolution of all chemical potentials.
The standard cosmic trajectory is obtained assuming tiny lepton and baryon asymmetries. 
For larger lepton asymmetry, the charge chemical potential exceeds the baryon chemical potential
before pion annihilation.
\end{abstract}

\pacs{95.30.Tg, 
11.30.Fs, 
05.70.Jk, 
12.38.Gc 
}
\keywords{cosmic QCD transition, lepton asymmetry}

\maketitle

The excess of matter over antimatter in the Universe is one of the 
major puzzles of particle physics and cosmology. 
This asymmetry can be specified with respect to the conserved charges: baryon number $B$, lepton number $L$, and electric charge $Q$.
The baryon asymmetry, defined as the net baryon number density per 
entropy density, $b=n_B/s$, is tightly constrained to be $b = (8.60 \pm 0.06) \times 10^{-11}$ inferred from \cite{Ade:2015xua}. 
However, the standard model of particle physics (SM) fails to explain this asymmetry \cite{Kolb:1983ni}.

The lepton asymmetry $l=n_L/s$ is a key parameter to understand the origin of the matter-antimatter asymmetry.
The idea of leptogenesis \cite{Buchmuller:2005eh} is to create a primeval lepton 
asymmetry which due to electroweak sphaleron processes is partly converted to a 
baryon asymmetry. In the SM, the prediction in the case of 
efficient sphaleron processes is $l=-(51/28)b$ \cite{Harvey:1990qw}. However, there exist also models that predict a large lepton asymmetry nowadays, i.e.,~$|l| \gg b$, and there is no preference for either sign of $l$.
In general, in these models sphaleron processes are either suppressed 
\cite{Harvey:1981cu, Casas:1997gx, Barenboim:2017dfq} or the lepton asymmetry is produced after sphaleron processes cease to be efficient \cite{Eijima:2017anv}. 

Observationally, the lepton asymmetry is only weakly constrained. While the charge neutrality of the Universe (see \cite{Caprini:2003gz} for an upper limit) links the asymmetry in charged leptons to the tiny baryon asymmetry, a much larger lepton asymmetry could reside in a large neutrino asymmetry today. Constraints on the lepton asymmetry can be obtained from the cosmic microwave background $\vert l \vert < 0.012$ (95\% C.L.) 
\cite{Oldengott:2017tzj} and are in concordance with big bang nucleosynthesis analyses \cite{Mangano:2011ip}.

In this Letter, we investigate how the cosmic trajectory through the QCD phase diagram is influenced by the unknown lepton asymmetry.
It has been shown by means of lattice QCD that the QCD transition is a crossover at small chemical 
potentials (see \cite{Aoki:2006we, Bhattacharya:2014ara} and e.g.\ \cite{Ding:2015ona} for a review). 
A pseudocritical temperature 
can be defined and is measured on the lattice to be $T_\mathrm{QCD} \simeq 154 (9)$ MeV \cite{Bazavov:2012jq} [$T_\mathrm{QCD} \simeq 147(2) - 165(5)$ MeV \cite{Borsanyi:2010bp}]. 
For vanishing temperatures and large baryon chemical potential $\mu_B>m_N$ effective models of QCD, like the Nambu--Jona-Lasinio model, predict a first-order chiral transition. It is speculated that there exists a critical line in the $(\mu_B, T)$ plane of the QCD phase diagram describing a first-order phase transition \citep{Asakawa:1989bq}. This line is expected to end in a second-order critical point \cite{Stephanov:1998dy, Stephanov:2004wx}.
Because of the infamous sign problem in lattice QCD, calculations for nonvanishing chemical potentials are very difficult \cite{Stephanov:2004wx, Philipsen:2007rm, Schmidt:2007jg, Ding:2017giu}:  
with nonzero chemical potentials, the Euclidean action
in the path integral for the partition sum becomes complex.
Then the exponential of the action is not positive definite. This invalidates 
a probabilistic interpretation of the path integral, and 
the usual numerical techniques of lattice QCD
are no longer applicable.

We present a novel technique of determining the evolution of chemical potentials for arbitrary lepton asymmetries throughout the QCD epoch by using lattice QCD susceptibilities. First calculations of the cosmic trajectory in the QCD phase diagram have been performed in \cite{Fromerth:2002wb}.
In \cite{Schwarz:2009ii} the evolution of chemical potentials at large lepton asymmetries has been studied in the approximation 
of an ideal quark gas and of a 
hadron resonance gas (HRG). 
In the HRG, the QCD-sector is approximated as an ideal gas of hadron resonances.
Here we extend and advance the approach of \cite{Schwarz:2009ii} in the following perspective: 
for the first time, we determine the cosmic trajectory accounting properly for strong interaction effects close to $T_\mathrm{QCD}$ by using lattice QCD data for conserved charge susceptibilities. 
In the context of sterile neutrino production, approximate relations between lepton asymmetries and chemical potentials have been studied in terms of susceptibilities in \cite{Ghiglieri:2015jua, Venumadhav:2015pla}.
Furthermore, we improve the calculation of the entropy density in \cite{Schwarz:2009ii} by including chemical potentials of all relevant particle species, and we take a larger number of hadron resonances into account.

The trajectory of the early Universe, for conserved $B,Q$, and $L$, in the phase diagram of strongly interacting matter is commonly assumed to pass $T_\mathrm{QCD}$ at vanishing chemical potentials and to proceed to $\mu_B=m_N$ and $\mu_Q\approx\mu_L\approx0$ at $m_e\ll T\lesssim m_N$. 
In this scenario, it is assumed that the lepton asymmetry is tiny, $l = \mathcal{O}(b)$. Below we refer to this as the standard scenario. Our results for $l=-(51/28)b$ present (to our knowledge) the first precise calculation of the standard cosmic trajectory.

It has already been shown that for $|l|\gg b$ the baryon and charge chemical potential becomes on the order of the lepton asymmetry $\mu_B \sim \mu_Q \sim lT$ at $T\gtrsim T_\mathrm{QCD}$ \cite{Zarembo:2000wj, Schwarz:2009ii}. 
Assuming an overall electric charge neutrality and a fixed $b$, a lepton asymmetry in the electrically charged leptons induces an electric charge asymmetry in the quark sector, which induces quark chemical potentials. As quarks carry not only electric charge but also baryon number, quark chemical potentials induce nonvanishing charge and baryon chemical potentials.
With a sufficiently large primordial lepton asymmetry, the cosmic trajectory could be shifted to higher charge and baryon chemical potential and thus the order of the QCD transition in the early Universe might be changed. 
This could have observable consequences via the production of
relics \cite{Schwarz:2003du,Witten:1984rs}, such as stochastic
gravitational waves, which could be measured by pulsar timing arrays \cite{Tiburzi:2018txc}.
Additionally, for charge chemical potential larger than the pion mass $m_{\pi}$, pion condensation might occur in the early Universe \cite{Abuki:2009hx, Brandt:2017oyy}. 
Understanding the impact of a lepton asymmetry on the evolution of the Universe at various epochs is therefore of crucial importance.

For the epoch of the cosmic QCD transition, kinetic and chemical equilibrium are excellent approximations. The timescales of interest are the interaction rates and the Hubble time, $t_H=1/H\simeq 10^{-5}$ s at $T_\mathrm{QCD}$, which is large compared to the timescales of strong, electromagnetic, and weak interactions. 

Since neutrino oscillations, which take place after the QCD epoch at $T_{\text{osc}}\sim 10$ MeV \cite{Mangano:2011ip, Dolgov:2002ab, Wong:2002fa}, lead to a mixing of all lepton flavors, the observational constraints hold for the total lepton asymmetry $l=\sum_{\alpha} l_{\alpha}$, $\alpha\in \{e,\mu,\tau\}$.
Thus, for $T>T_{\text{osc}}$ even larger but oppositely signed lepton flavor asymmetries are consistent with observational constraints. 
Note that sizable flavor asymmetries survive neutrino oscillations dependent on the neutrino mixing angles and the initial values of the lepton asymmetries and thus even after neutrino oscillations it is possible to have $l_{\alpha}\neq l/3$ \cite{Pastor:2008ti,Barenboim:2016shh,Johns:2016enc}. In this Letter we focus on equally distributed lepton flavor asymmetries $l_e=l_{\mu}=l_{\tau} = l /3$.

After the electroweak transition at $T_{\text{EW}}\sim 100$ GeV and
for $T>T_{\text{osc}}$, there are five conserved charges in the early
Universe, $B, Q$, and the three lepton flavor numbers $L_{\alpha}$, to
which corresponding chemical potentials $\mu_B$, $\mu_Q$, and
$\mu_{L_{\alpha}}$ can be assigned, respectively.  Note that, in
contrast to relativistic heavy ion collisions that also probe the QCD
phase diagram, individual quark flavors like strangeness are not
conserved due to electroweak processes.

Assuming a homogeneous Universe, we obtain five conserved quantities:
$n_{L_{\alpha}}/s=l_{\alpha}$, $n_{B}/s=b$, and $n_{Q}/s=q$. The
entropy density $s$ fulfills the relation $Ts(T,\mu)=\epsilon(T,\mu) +
p(T,\mu)-\sum_a\mu_an_a(T,\mu)$, with $\epsilon$ the total energy
density, $p$ the total pressure, and the sum over conserved charges
$a\in \lbrace B,Q,L_{\alpha} \rbrace$. We fix $b = 8.6 \times
10^{-11}$ and $q=0$ in agreement with observations.  The three lepton
flavor asymmetries $l_{\alpha}$ remain free parameters.
 
The net number density of a particle species is defined as the number
density of a particle minus the density of its antiparticle. Assuming
chemical equilibrium, we find relations between the chemical potentials
of different particle species. Since photons and gluons carry no
conserved charges, their chemical potentials are zero. It follows that
the chemical potentials of particles and antiparticles are equal in
magnitude and opposite in sign, i.e., $\mu_i=-\mu_{\bar{i}}$. In
kinetic and chemical equilibrium for an ideal gas we can express the
net number densities by the integral over the Fermi-Dirac
(Bose-Einstein) distribution for fermions (bosons) as
\begin{eqnarray}
  n_i
  =&&
  \frac{g_i}{2\pi^2}\int_{m_i}^{\infty}\mathrm d E E\sqrt{E^2-m_i^2}
  \nonumber \\
  &&
\times\left( \frac{1}{e^{(E-\mu_i)/T}\pm 1}-\frac{1}{e^{(E+\mu_i)/T}\pm 1}\right)\ ,
\label{eq:netnumb}
\end{eqnarray}
for a particle with mass $m_i$, chemical potential $\mu_i$, and with the $+$ for fermions ($-$ for bosons). The number of degrees of freedom $g_i$ counts particles and antiparticles 
separately, i.e., $g=1$ for neutrinos, $g=2$ for electrically charged leptons, and 
$g=6$ for quarks. 
In this approximation, the five local conservation laws can be written in terms of the particle net number densities,
\begin{subequations}
\begin{eqnarray}
\label{eq:ls}
l_{\alpha}s&&=n_{\alpha}+n_{\nu_{\alpha}}\ , \\
\label{eq:bs}
bs&&=\sum_i B_in_i\ ,\\
\label{eq:qs}
qs&&=\sum_i Q_i n_i\  , 
\end{eqnarray}
\end{subequations}
with $B_i$ the baryon number and $Q_i$ the electric charge of particle species $i$.

We can express the conserved charge chemical potentials in terms of particle chemical potentials,
\begin{eqnarray}
\label{eq:muLa}
\mu_{L_\alpha}&&=\mu_{\nu_{\alpha}}\ , \\
\label{eq:muQLa}
\mu_{Q}&&=\mu_{\nu_{\alpha}}-\mu_{\alpha}=\mu_u-\mu_d\ ,\\ 
\mu_B&&=\mu_u+2\mu_d\ ,
\end{eqnarray}
and at low temperatures
\begin{eqnarray}
\mu_{Q}&&=\mu_{\pi}=\mu_p-\mu_n\ , \\
\label{eq:muBmun}
\mu_B&&=\mu_n\ ,
\end{eqnarray}
with $\mu_{\pi}$, $\mu_p$, and $\mu_n$ the chemical potential of pions, protons, and neutrons, respectively, and similar for other hadrons and their resonances. 

In order to take into account strong interactions between quarks and gluons close to $T_{\mathrm{QCD}}$
we expand the QCD pressure in a 
Taylor series in the chemical potentials up to second order,
\begin{eqnarray}
\label{eq:pressureQCD}
p^\mathrm{QCD}\!(T,\mu)\!=\!p^{\mathrm{QCD}}\!(T,0)+\frac{1}{2}\mu_a\chi_{ab}(T)\mu_b +\mathcal{O}(\mu^4)\ ,
\end{eqnarray}
with an implicit sum over $a,b\in \{B,Q\}$ here and in the following. The susceptibilities are defined by 
\begin{eqnarray}
\chi_{ab}(T)=\frac{\partial^2p^{\mathrm{QCD}}(T,\mu)}{\partial\mu_a\partial\mu_b}\bigg\vert_{\mu=0}=\chi_{ba}(T)\ .
\end{eqnarray}
Such an expansion is also used in lattice QCD for circumventing the sign problem (cf. \cite{Gavai:2001fr, Bazavov:2012jq}).
The conserved charge densities follow as
\begin{eqnarray}
\label{eq:netnumbQCD}
n_a(T,\mu)=\frac{\partial p^{\mathrm{QCD}}(T,\mu)}{\partial \mu_a}=\chi_{ab}\mu_b+ \mathcal{O}(\mu^3) \ .
\end{eqnarray}
The entropy density of the strongly interacting matter satisfies (cf. \cite{Bazavov:2017})
\begin{eqnarray}
Ts^{\mathrm{QCD}}(T,\mu)=T \frac{\partial p^{\mathrm{QCD}}}{\partial T} - \sum_a\mu_a\frac{\partial p^{\mathrm{QCD}}}{\partial \mu_a}\ ,
\end{eqnarray}
and with  Eqs.~\eqref{eq:pressureQCD} and \eqref{eq:netnumbQCD} it follows that
\begin{eqnarray}
\label{eq:entropyQCD}
s^{\mathrm{QCD}}\!(T,\mu)\! -\! s^{\mathrm{QCD}}\!(T,0)\!=\!\left(\!\frac{1}{2}\!\frac{\mathrm d \chi_{ab}}{\mathrm d T}\!-\!\frac{1}{T}\chi_{ab}\!\right)\!\mu_a\mu_b \ .
\end{eqnarray}

Only quarks contribute to the baryon asymmetry $bs=n_B^{\mathrm{QCD}}$. The contribution to the electric charge can be divided into a part arising from the leptons $n_Q^{\mathrm{lep}}$ and one by quarks $n_Q^{\mathrm{QCD}}$:
$qs=0=n_Q^{\mathrm{QCD}}+n_Q^{\mathrm{lep}}$. With the QCD net number densities given by Eq.~\eqref{eq:netnumbQCD}, our system of equations becomes
\begin{subequations}
\begin{eqnarray}
\label{eq:lsQCD}
l_{\alpha}s&&=n_{\alpha}+n_{\nu_{\alpha}}\ , \\
\label{eq:bsQCD}
bs&&=\mu_B\chi_{BB} + \mu_Q\chi_{BQ}\ ,\\
\label{eq:qsQCD}
qs&&=\mu_Q\chi_{QQ} + \mu_B\chi_{BQ}-\sum_{\alpha}n_{\alpha}\  .
\end{eqnarray}
\end{subequations}

For given temperature and lepton asymmetry $l$, we solve Eqs.~\eqref{eq:ls}--\eqref{eq:qs} or, respectively, Eqs.~\eqref{eq:lsQCD}--\eqref{eq:qsQCD} for 
(i) the ideal quark gas ($T\geq 100$ MeV), 
(ii) lattice QCD susceptibilities ($250$ $\geq T\geq 150$ MeV), 
and (iii) HRG ($250$ $\geq T\geq 10$ MeV).

In order to solve the system of coupled integral equations we modified the \texttt{C} code used in \cite{Schwarz:2009ii} to take into account strong interactions between quarks according to 
Eqs.~\eqref{eq:entropyQCD} and \eqref{eq:lsQCD}--\eqref{eq:qsQCD}. We use continuum extrapolated lattice QCD susceptibilities for a 2+1 flavor system \cite{Bazavov:2012jq}, i.e., including the up, down, and strange quark, and for a 2+1+1 flavor system (not continuum extrapolated, $N_{\tau}=8$) \cite{Bazavov:2014yba, Mukherjee:2015mxc}, i.e.,\ including also the charm quark, and the numerical temperature derivatives thereof. 
For the HRG we consider hadron resonances up to $m_{\Lambda(2350)}\approx 2350 $ MeV $\sim 15T_{\mathrm{QCD}}$, using particle properties  
according to the summary tables in \cite{PDG:2018}. 
Finally, we included chemical potentials in the calculation of the entropy density in all three temperature regimes (i)-(iii) and took $s^{\mathrm{QCD}}\!(T,0)$ by \cite{Laine:2006cp}. 
Integrations like in Eq.~\eqref{eq:netnumb} are performed using Gauss-Laguerre quadrature and the system of equations is solved by using Broydn's method \cite{NumericalRecipes}. Deviations from 
the results of \cite{Schwarz:2009ii} are due to those improvements and minor mistakes in the original 
code.
We are free to choose arbitrarily five independent chemical potentials as free parameters according to Eqs.~\eqref{eq:muLa}--\eqref{eq:muBmun} in order to solve our system of integral equations. However, one has to carefully choose them such that they are of different size to be able to obtain all particle chemical potentials without running into numerical problems. This is most important for the HRG at low temperatures where $\mu_Q=\mu_p-\mu_n$ and $\mu_n\approx\mu_p$. A good choice is $\lbrace \mu_Q,\mu_B, \mu_{L_e}, \mu_{L_{\mu}}, \mu_{L_{\tau}}\rbrace$.

\begin{figure}
\includegraphics[width=\columnwidth]{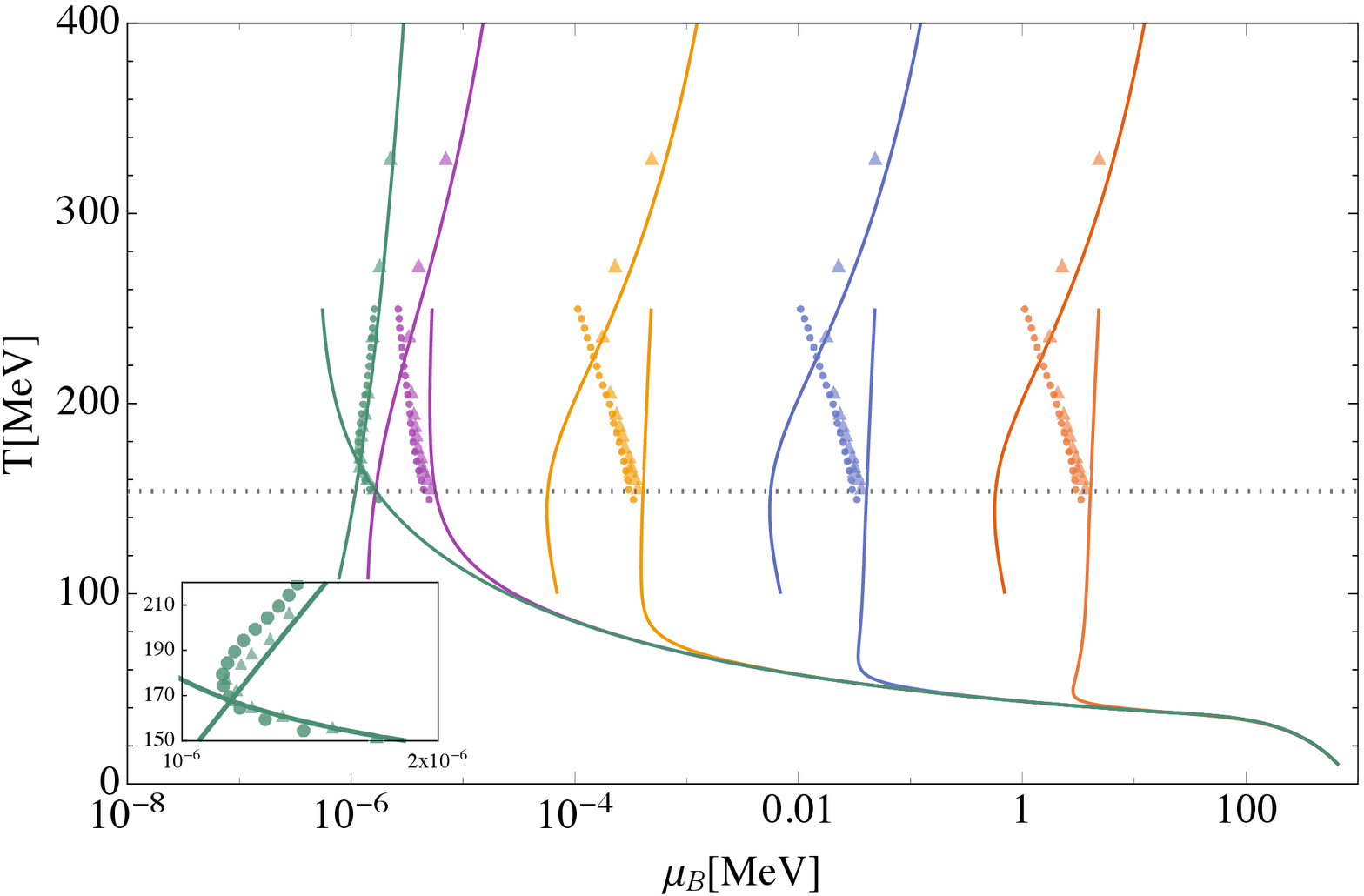}

\vspace{0.5cm}
\includegraphics[width=\columnwidth]{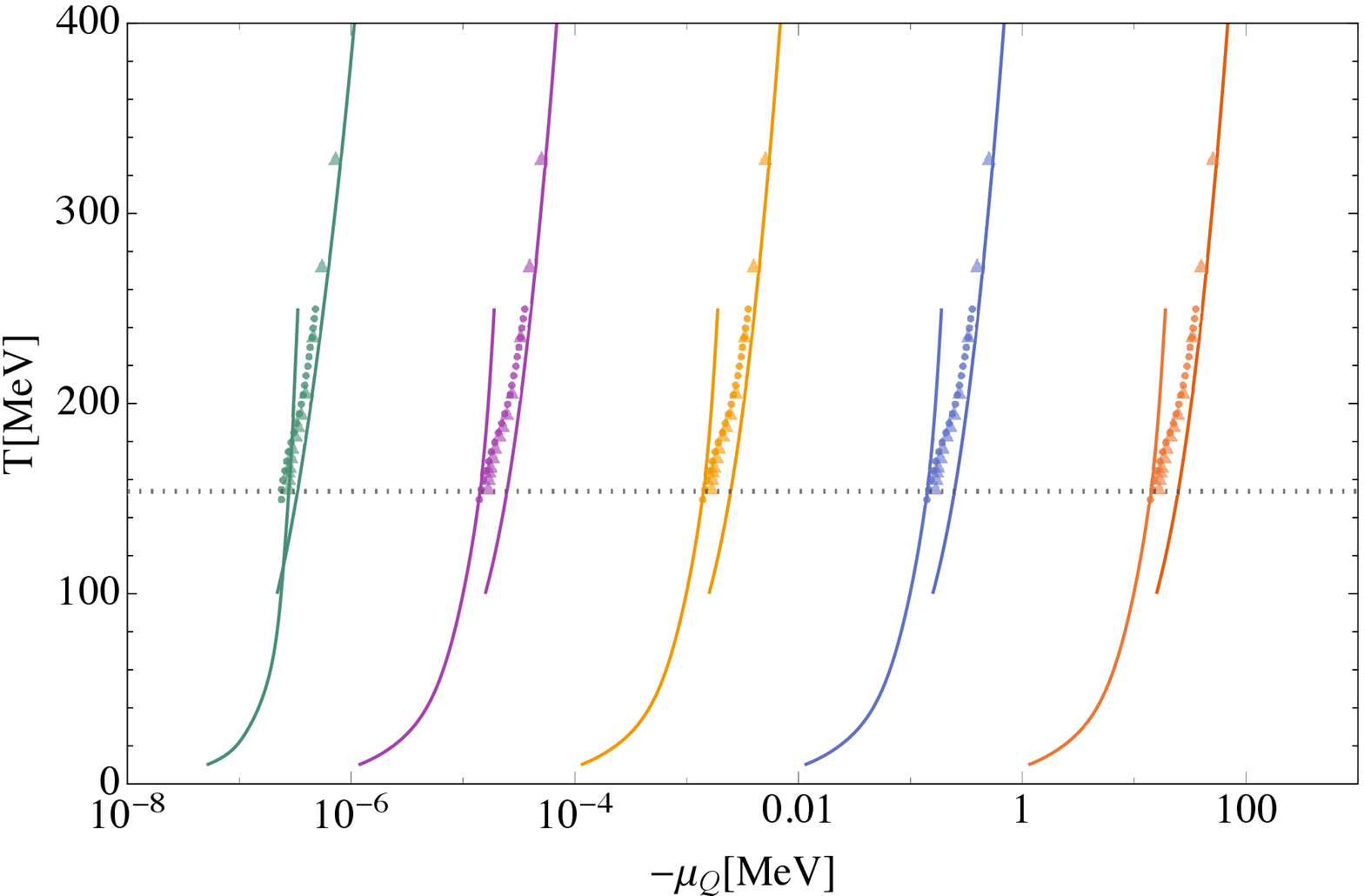}

\vspace{0.5cm}
\includegraphics[width=\columnwidth]{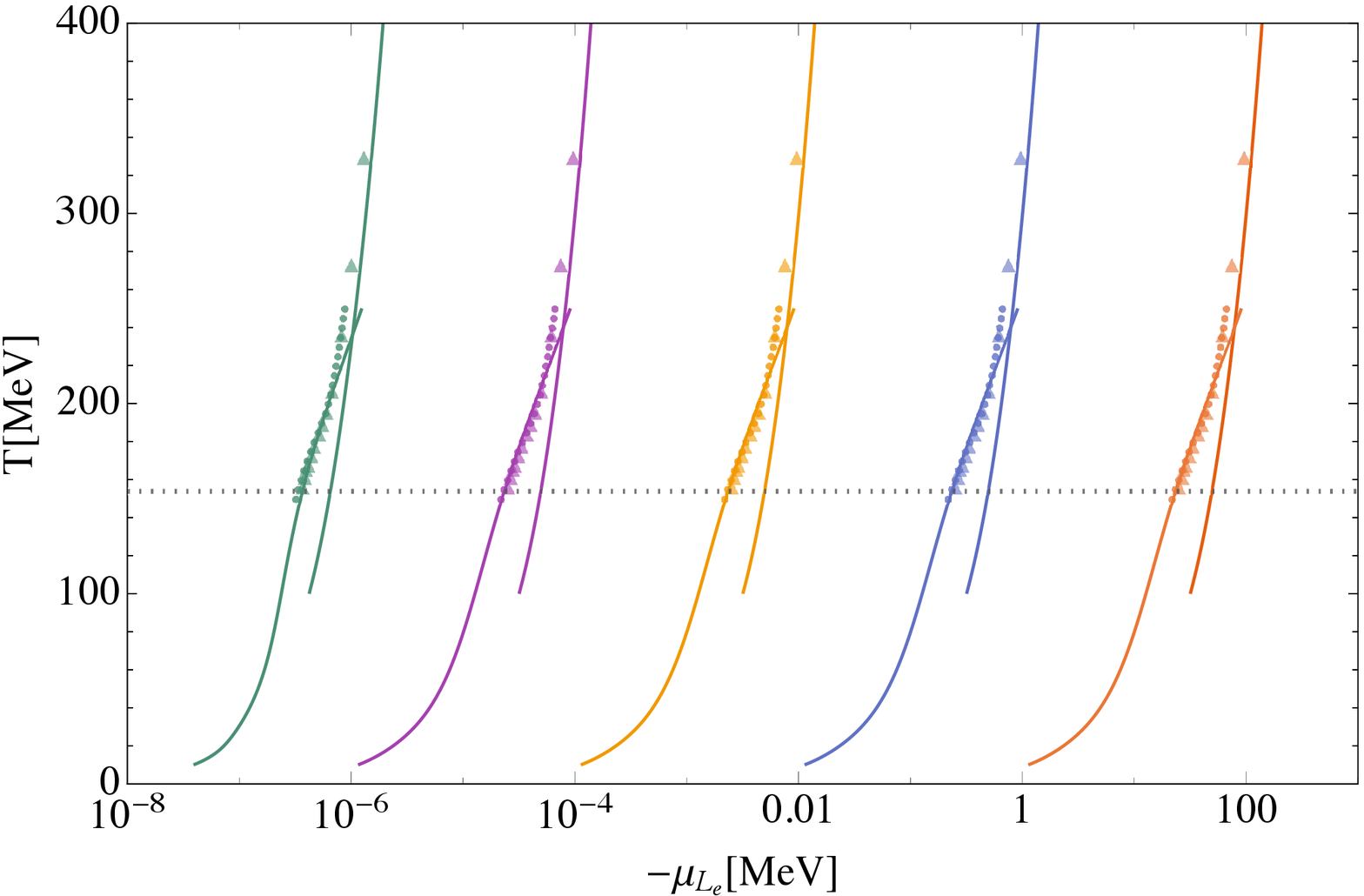}

\vspace{0.5cm}
\includegraphics[width=\columnwidth]{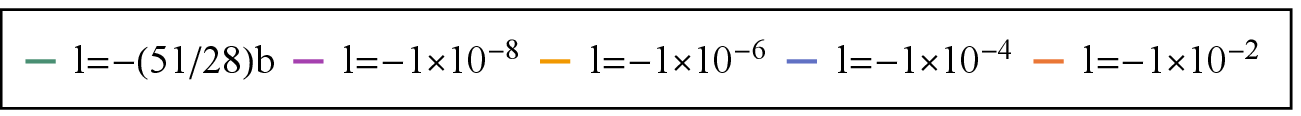}
\caption{\label{fig:muT}Temperature evolution of chemical potentials for different negative total lepton asymmetries $l$. (Top) Baryon chemical potential $\mu_B$. (Middle) Electric charge chemical potential $-\mu_Q$. (Bottom) Electron lepton flavor chemical potential $-\mu_{L_e}$. Continuous lines for high temperatures are results for the ideal quark gas, for low temperatures for the HRG. The symbols 
$\bullet$ and $\blacktriangle$ indicate results obtained by using 2+1 and 2+1+1 flavor lattice QCD susceptibilities, respectively.
The magnitude of the lepton asymmetry increases from left to right. The pseudocritical temperature $T_{\mathrm{QCD}}\approx 154 $ MeV is displayed by a horizontal dotted line. The standard cosmic trajectory of the early Universe is given by $l=-(51/28)b$.}
\end{figure}

Figure~\ref{fig:muT} shows the results for the temperature evolution of $\mu_B$ (top), $-\mu_Q$ (middle), 
and $-\mu_{L_e}$ (bottom) for different values of the lepton asymmetry. The evolution of $\mu_{L_{\mu}}$ and $\mu_{L_{\tau}}$ is not shown here, 
as they are of similar size as $\mu_{L_{e}}$. 
However, despite the fact that all lepton asymmetries $l_\alpha$ are assumed to be equal, the 
three lepton flavor chemical potentials evolve 
differently due to the lepton masses.
We can see in Fig.~\ref{fig:muT} that, talking about absolute values, a larger total lepton 
asymmetry induces larger chemical potentials. The chemical potentials are proportional to $l$.
This is true for $l > \mathcal O (b)$. For lepton asymmetries $l\lesssim\mathcal O(b)$ the evolution of all chemical potentials is determined by the baryon asymmetry $b$ and the contribution of $l$ is negligible. 

\begin{figure}
\includegraphics[width=\columnwidth]{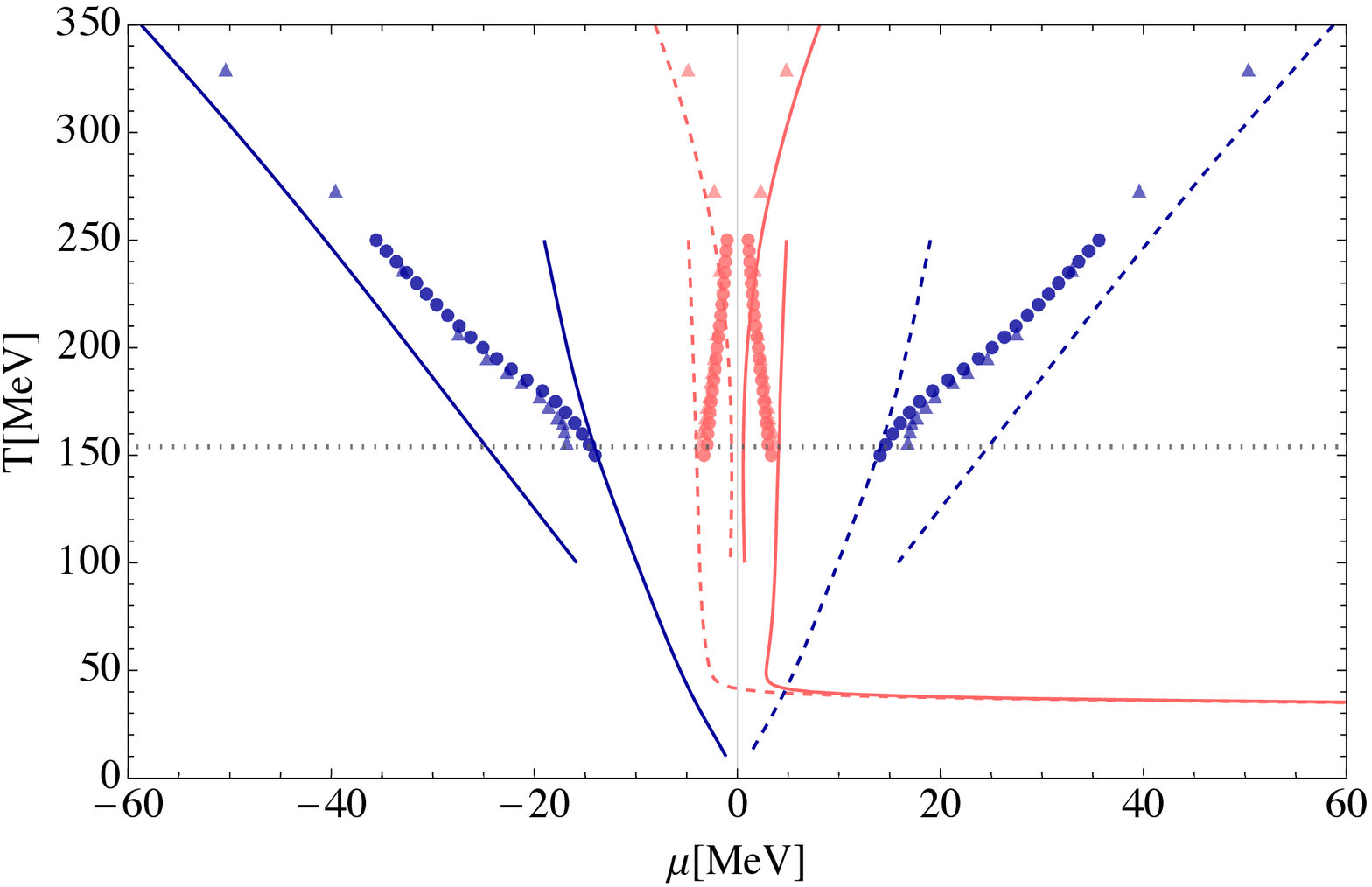}
\includegraphics[scale=0.65]{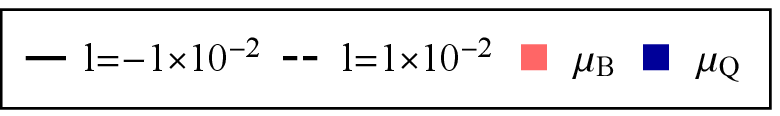}
\caption{\label{fig:muBzoom}Temperature evolution of baryon chemical potential $\mu_B$ 
and electric charge chemical potential $\mu_Q$ for both
signs of a large lepton asymmetry.
Continuous and dashed lines for high temperatures are results for the ideal quark gas, for low 
temperatures for the HRG. The symbols $\bullet$ and $\blacktriangle$ indicate 
results obtained by using 2+1 and  2+1+1 flavor lattice QCD susceptibilities, respectively.}
\end{figure}

The chemical potentials obtained using lattice QCD susceptibilities connect the ideal quark gas with the HRG approximation quite well. Especially for $\mu_Q$ they almost smoothly connect the two approximations at high and low temperature. For $\mu_B$, however, the results with 2+1 flavor lattice QCD susceptibilities do not connect the two approximations smoothly for larger lepton asymmetries.
At low temperatures there is a small gap between the lattice QCD and HRG results.
At high temperatures the lattice QCD results do not smoothly converge to the ideal quark gas, but they intersect in a single point for $\vert l\vert \gtrsim 10^{-8}$. Taking into account the charm quark by 
using  2+1+1 flavor lattice QCD susceptibilities, 
they seem to converge to the ideal quark gas at high temperatures [see Fig.~\ref{fig:muT} (top)]. The uncertainty of the lattice QCD results is on the order of the point sizes in Figs.~\ref{fig:muT} and \ref{fig:muBzoom}.

An important feature in the evolution of $\mu_B$ is that, for small temperatures $T\lesssim m_{\pi}/3\approx 46$ MeV, after the annihilation of pions (and muons), $\mu_B$ no longer depends
on the value of $l$ and approaches the nucleon mass $m_N\sim 1 $ GeV at low temperatures (see Fig.~\ref{fig:muT}). 

For $l=-(51/28)b$ we obtain the standard cosmic trajectory. The reader should keep in mind that the cosmic trajectory follows a path in the $5+1$-dimensional phase diagram.
In Fig.~\ref{fig:muT} we show two-dimensional projections of the phase diagram.

It can also be seen that, for large lepton asymmetries $l>\mathcal O (b)$ (see Figs.~\ref{fig:muT} and \ref{fig:muBzoom}), the electric charge 
chemical potential becomes larger than the baryon chemical potential 
at nonvanishing temperature. This can be understood as follows.
The electric charges of the three light quarks add up to zero. 
If their masses were degenerate (and heavier quarks are neglected),
the  susceptibility $ \chi  _ { BQ } $ would vanish, so that
no $\mu_B $ is induced. Thus, for $T \gtrsim m_{\rm strange}$,
 $ \mu_B $ remains small. Furthermore, this is why the
charm quark is important here, despite its large mass. 
 
Figure~\ref{fig:muBzoom} shows the effect of 
the sign of a lepton asymmetry.
For positive $l$, $\mu_B$ is negative at high temperatures and proceeds to the nucleon mass for lower temperatures. 

For an equally distributed lepton asymmetry, we find that for $\vert l\vert \gtrsim 0.15 $ we get $\vert \mu_Q\vert\gtrsim m_{\pi}$, which  might enable pion condensation in the early Universe \cite{Abuki:2009hx, Brandt:2017oyy}. Such a large lepton asymmetry would exceed the observational constraint
by an order of magnitude. However, unequally distributed lepton asymmetries would admit 
the possibility of $\vert\mu_Q\vert\gtrsim m_{\pi}$, while satisfying $\vert l \vert <0.012$.

In this Letter, we have studied the evolution of chemical potentials as a function of temperature during the cosmic QCD epoch and investigated its dependence on a lepton asymmetry. For the first time, we used lattice QCD results to properly account for the temperature regime around $T_{\mathrm{QCD}}$ in order to connect the approximations of an ideal quark gas with the HRG. We provide the standard cosmic trajectory through the 5+1-dimensional QCD phase diagram for $l=-(51/28)b$ and, furthermore, the cosmic trajectory 
in the presence of larger total lepton asymmetry.

There is no phase transition in the early Universe if the cosmic trajectory 
in the QCD phase diagram is smooth at all $T$. A kink in the trajectory would correspond to a 
nonequilibrium, first-order phase transition. We find that the 2+1+1 flavor lattice QCD susceptibilities allow us to interpolate between the trajectories of the ideal quark and HRG. We like to stress the importance of the charm quark contribution in our results to obtain a smooth trajectory. 
Unfortunately, no continuum extrapolated 2+1+1 flavor lattice QCD susceptibilities were
available at the time of this study.

Gaps in our result for the cosmic trajectory, like in Fig. \ref{fig:muBzoom}, are artifacts of our approximations. 
Gaps between the lattice QCD results and the ideal quark gas might be closed by taking higher-order perturbative corrections into account. 
It has been shown that these lead to smaller susceptibilities than in the ideal gas approximation and to better agreement with lattice QCD results \cite{Vuorinen:2002ue}. 
Furthermore, it would be helpful to have lattice QCD susceptibilities for lower and higher temperatures available. 
The observed small gaps for $\mu_B$ at low temperature between lattice QCD results and the HRG approximation might then be closed. However, these gaps might also be due to limitations of the HRG approximation to describe all thermodynamical aspects of QCD.
The current precision of lattice susceptibilities and the ideal quark gas and HRG approximations 
used in this Letter do not allow us to make any statement on the nature of the cosmic QCD transition. 

Our results might be of crucial importance for a better understanding of the evolution of the early 
Universe and can be used in cosmic evolution calculations, e.g.,\ in predicting the abundance of various dark matter candidates. 
Furthermore, our framework can be easily extended to study the influence of additional particles beyond the SM or resonances that are in kinetic and chemical equilibrium with the SM particles. If the interactions of the new particles violate some of the charge conservation, further modifications of our framework are necessary. 
We would like to emphasize that, before pion annihilation and for a lepton asymmetry $\vert l\vert
\gtrsim 10^{-8}$, the absolute value of the electric charge chemical potential $\vert \mu_Q \vert$
exceeds the baryon chemical potential and therefore $\mu_Q$ might be more important for the 
thermal history of the Universe than $\mu_B$.

\begin{acknowledgments}
We thank Gabriela Barenboim, Frithjof Karsch, Pascal Kreling, 
J\"urgen Schaffner-Bielich, Christian Schmidt, Maik Stuke, and Joris Verbiest for fruitful discussions.
M. M. W. acknowledges the support by Studienstiftung des Deutschen Volkes. 
I. M. O. acknowledges support from the Ministerio
de Educación y Ciencia (MEC) and FEDER (EC) Grants No. SEV-2014-0398, FIS2015-2245-EXP, and FPA2014-54459 and the Generalitat Valenciana under Grant No. PROME-TEOII/2013/017.
All authors acknowledge support by the Deutsche Forschungsgemeinschaft (DFG) through the Grant No. CRC-TR 211 ``Strong-interaction matter under extreme conditions''.
\end{acknowledgments}

\bibliography{CosmicQCDepoch_v2.bib}

\end{document}